\documentclass[sigconf,screen]{acmart}

\AtBeginDocument{%
  }

\copyrightyear{2025}
\acmYear{2025}
\acmMonth{March}
\usepackage{xspace}
\usepackage{enumitem}
\usepackage{tabularx}
\usepackage{multicol}
\usepackage{multirow}
\usepackage{amsmath}
\usepackage{graphicx}
\usepackage{algorithmic}
\usepackage{algorithm}
\usepackage{booktabs}
\usepackage{makecell}
\usepackage{caption}
\usepackage{colortbl}
\usepackage{tikz}
\definecolor{mytitle}{rgb}{0.9, 0.9, 0.9}
\definecolor{mygray}{rgb}{0.95, 0.95, 0.95}

\renewcommand{\refname}{REFERENCES}

\newcommand{\toolname}{\textsc{PriceSleuth}\xspace}

\setcopyright{acmlicensed}\acmConference[Internetware'25 - New Idea Track]{2025 16th International Conference on Internetware: New Idea Track}{June 20-22, 2025}{Trondheim, Norway}
\acmBooktitle{16th International Conference on Internetware: New Idea Track, June 20-22, 2025, Trondheim, Norway}

\begin{document}

\title{Detecting State Manipulation Vulnerabilities in Smart Contracts Using LLM and Static Analysis}



\author{Hao Wu}
\email{emmanuel_wh@stu.xjtu.edu.cn}
\affiliation{\institution{Xi'an Jiaotong University}\city{Xi'an}\country{China}}

\author{Haijun Wang}
\email{haijunwang@xjtu.edu.cn}
\authornote{Corresponding author}
\affiliation{\institution{Xi'an Jiaotong University}\city{Xi'an}\country{China}}

\author{Shangwang Li}
\email{3124357023@stu.xjtu.edu.cn}
\affiliation{\institution{Xi'an Jiaotong University}\city{Xi'an}\country{China}}

\author{Yin Wu}
\email{wuyin@stu.xjtu.edu.cn}
\affiliation{\institution{Xi'an Jiaotong University}\city{Xi'an}\country{China}}

\author{Ming Fan}
\email{mingfan@mail.xjtu.edu.cn}
\affiliation{\institution{Xi'an Jiaotong University}\city{Xi'an}\country{China}}

\author{Yitao Zhao}
\email{zyt717@hotmail.com}
\affiliation{\institution{Yunnan Power Grid Co., Ltd.}\city{Yunnan}\country{China}}

\author{Ting Liu}
\email{tingliu@mail.xjtu.edu.cn}
\affiliation{\institution{Xi'an Jiaotong University}\city{Xi'an}\country{China}}

\renewcommand{\shortauthors}{Hao Wu, Haijun Wang, Shangwang Li, Yin Wu, Ming Fan, Yitao Zhao, and Ting Liu}

\begin{abstract}
An increasing number of DeFi protocols are gaining popularity, facilitating transactions among multiple anonymous users. 
State Manipulation is one of the notorious attacks in DeFi smart contracts, with price variable being the most commonly exploited state variable—attackers manipulate token prices to gain illicit profits.
In this paper, we propose \toolname, a novel method that leverages the Large Language Model (LLM) and static analysis to detect Price Manipulation (PM) attacks proactively. \toolname firstly identifies core logic function related to price calculation in DeFi contracts. Then it guides LLM to locate the price calculation code statements. Secondly, \toolname performs backward dependency analysis of price variables, instructing LLM in detecting potential price manipulation. Finally, \toolname utilizes propagation analysis of price variables to assist LLM in detecting whether these variables are maliciously exploited.
We presented preliminary experimental results to substantiate the effectiveness of \toolname. And we outline future research directions for \toolname.

\end{abstract}






\maketitle

\section{Introduction}
Smart contracts are programs, which are decentralized, anonymous, and self-executing without relying on untrusted third parties. Due to these characteristics, smart contracts enable users to participate in various finance activities, facilitating the advancement of decentralized finance (DeFi). DeFi protocols are made up of a series of smart contracts deployed on the blockchain to manage digital assets. However, given that the smart contracts are publicly accessible on the blockchain, potential vulnerabilities within DeFi protocols can be easily exploited to yield unfair profits. A prevalent form of exploitation is the State Manipulation attack~\cite{zhang2023demystifying}. Among various state manipulation strategies, Price Manipulation (PM) has emerged as one of the most financially devastating forms, which targets the economic parameters that are most sensitive in DeFi ecosystems.

Existing methods for detecting PM attacks can be categorized into transaction-based methods~\cite{DeFiRanger, wang2024defiguard} and code-based methods~\cite{kong2023defitainter,zhang2025following}. Transaction-based methods~\cite{DeFiRanger, wang2024defiguard} typically analyze transaction behaviors, employing pattern recognition to detect such attacks. However, these methods can only identify attacks post hoc, limiting their ability to provide timely detection and effective mitigation. In contrast, code-based methods~\cite{kong2023defitainter,zhang2025following} rely on predefined price manipulation patterns and leverage static analysis techniques for detection. Yet, real-world price manipulation attacks are complex, making it challenging to establish comprehensive rules. Moreover, static analysis often struggles to accurately reason the semantics of DeFi operations, leading to high false positives and false negatives. 

In real-world scenarios, DeFi protocols are complex, and attackers need to coordinate multiple behaviors to execute a successful attack. Consequently, they typically use smart contracts, known as attack contracts, to implement the process of exploiting vulnerabilities. To address the timeliness of detection, we aim to detect PM attack contracts before the execution of the attack, providing early warnings and proactive defense. Furthermore, to tackle the evolving complexity nature of price manipulation and the semantic limitations of static analysis, we combine static analysis and Large Language Models (LLMs)~\cite{sun2024gptscan} to understand the semantics of price calculation mechanisms (PCMs), further identifying manipulable PCMs to detect PM attacks.

In this paper, we introduce \toolname, a real-time detection framework that combines LLMs with static analysis to identify PM attack contracts. We first employ static analysis to identify logic functions related to PCMs in DeFi contracts. Then we leverage the reasoning capabilities of LLMs to locate the PCMs code. Subsequently, we perform backward data dependency analysis of price variables, instructing LLM to detect whether the PCMs are susceptible to manipulation. Finally, we utilize propagation analysis of the manipulated price values to assist LLMs in verifying whether the manipulated price ultimately leads to malicious exploitation. We conducted preliminary experiments on four types of DeFi protocols to validate the effectiveness of \toolname. The results demonstrate that although different types of DeFi protocols have different code logic, \toolname is capable of understanding the semantics of PCMs and detecting the four PM attacks. And \toolname successfully detects PM attack contracts within one minute, making it a practical solution for proactive attack detection.


\section{Motivating Example}

\begin{figure}[h]
    \centering
    \includegraphics[width=0.96\linewidth]{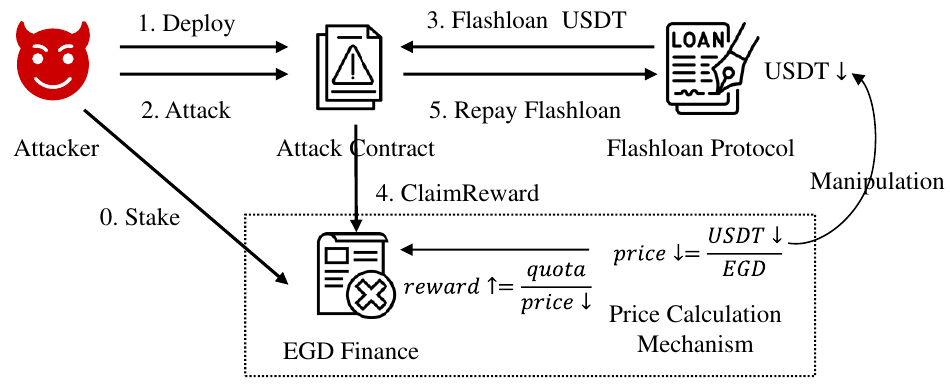}
    \caption{A Simplified Attack Flow against EGD Finance}
    \label{fig:motivating example}
    \vspace{-0.2cm}
\end{figure}

As shown in Figure~\ref{fig:motivating example}, we provided an example of PM attack to illustrate our motivation. The example is from a real-world PM attack on EGD Finance on Augest 2022, which caused a financial loss of \$36K ~\cite{EGDFinance}. 
EGD Finance, a yield farming DeFi protocol~\cite{zhong2025defiscope}, allows users to stake cryptocurrencies through the \textit{stake()} function and later withdraw their principal and earned rewards via the \textit{ClaimReward()} function.
The attack consists of five key steps, with the root cause stemming from the price calculation mechanisms that are susceptible to manipulation.

The attackers began by calling the \textit{stake()} function to deposit USDT into EGD Finance as a preparatory step. Next, they encoded the attack logic into a attack contract and deployed it on the chain. The attackers then invoked the contract. It initiated the exploit by leveraging a flashloan protocol (i.e., PancakeSwap) to borrow a large amount of \textbf{USDT}. Following the flashloan, the attack contract called the \textit{ClaimReward()} function in EGD Finance to claim rewards. However, the reward amount is directly tied to the price of the EGD token, which is determined by the ratio of the amount of \textbf{USDT} to \textbf{EGD} in the PancakeSwap liquidity pool. The attackers identified this price calculation mechanism is manipulable and used the flashloan to reduce the amount of \textbf{USDT} in the liquidity pool, causing the \textbf{EGD} token price to skyrocket artificially. As a result, the attackers were able to withdraw an excessive amount of rewards.

In the EGD Finance attack incident, we observed a 6-minute interval between the deployment of the attack contract and the initiation of the attack. If the attack contract could be identified during this period, appropriate mitigation measures could be taken to prevent financial losses, which has been validated in previous studies~\cite{wang2024skyeye, zhang2025following}. Furthermore, we observed that the root cause of PM attack lies in easily manipulated price calculation mechanisms. Traditional program analysis techniques struggle to accurately reason the semantics of code. Based on above observations, we propose a novel PM attack contract detection approach integrating LLMs and static analysis to enhance the capability of PM attack detection.

\section{Methodology}

\begin{figure}[h]
    \centering
    \includegraphics[width=0.96\linewidth]{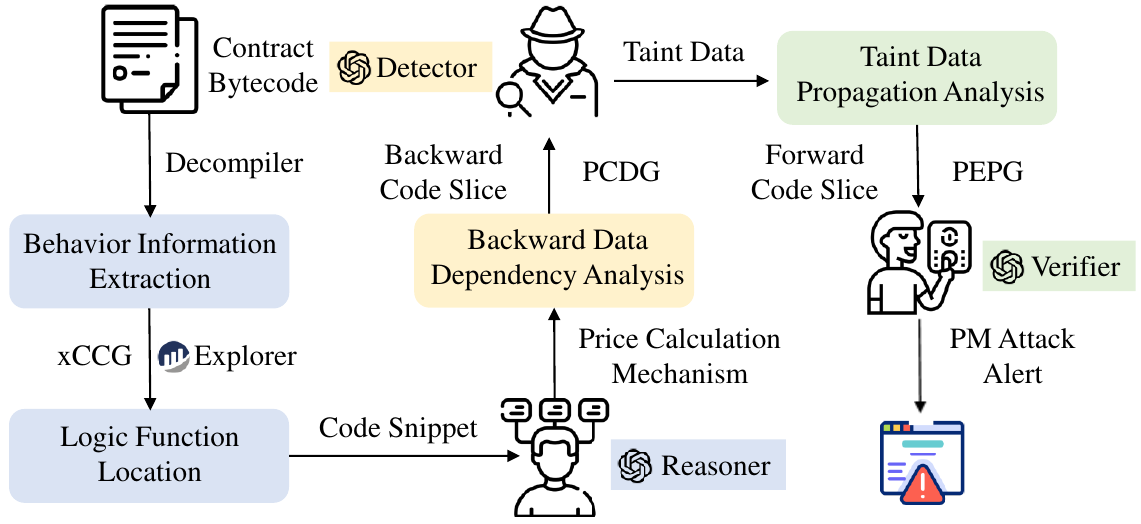}
    \caption{Overview of \toolname}
    \label{fig:overview}
    \vspace{-0.25cm}
\end{figure}

Figure~\ref{fig:overview} shows the overview design of \toolname, consisting of three main modules. Initially, given the bytecode of a smart contract, \toolname leverages the open-source decompiler~\cite{grech2019gigahorse} to decompile the contract bytecode and extract behavior information. Then it constructs the inter-contract call graph (xCCG). By leveraging the xCCG and explorer (e.g.,Etherscan), \toolname locates the logic function, which is used by the \textbf{Reasoner LLM} to extract the price calculation mechanism (PCM). The second step involves backward data dependency analysis to trace the price variable dependencies, generating price calculation dependency graphs (PCDG). These code snippets and PCDG are used by the \textbf{Detector LLM} to analyze whether PCM is vulnerable to manipulation. In the final step, \toolname employs propagation analysis to trace the tainted price data, generating the price exploitation propagation graphs (PEPG). Then \textbf{Verifier LLM} leverages the PEPG and the corresponding code snippet to reason about the code and verify whether the manipulated price leads to malicious exploitation.

\subsection{Price Calculation Mechanism Reasoning}
In this section, we introduce how \toolname extracts the price calculation mechanisms involved in the DeFi contracts.

\subsubsection{Behavior Information Extraction}
Attackers often conceal their malicious intent by keeping their deployed contract code closed-source. Therefore, we employ the decompiler Gigahorse~\cite{grech2019gigahorse} to decompile the contract bytecode. To obtain external call information, we need to identify the callee contract address and function. For addresses hard-coded within the contract, we use data-flow analysis to extract the constant addresses. For addresses stored in storage, we utilize the Web3 API \textit{getStorageAt} combined with the slot number and offsets to locate the storage addresses. In other cases where the callee address is dynamically provided through \textit{Calldata}, we adopt the probabilistic matching approach proposed in~\cite{wang2024skyeye} to determine the most likely invoked contract address. Then we extract the callee function signatures by data-flow analysis.

\subsubsection{Logic Function Location} \label{code snippet}
After obtaining the behavior information of the analyzed contract, we construct the inter-contract call graph (xCCG). To enhance analysis efficiency, we prioritize the traversal of the xCCG. Since PM attacks typically require a large amount of assets, we hypothesize that attackers are likely to leverage flashloan as a funding mechanism. Therefore, we assign higher priority to branches that involve flashloan callback function signatures.
Following the assigned priority order, we traverse the xCCG, retrieving the DeFi contract code via blockchain explorer API. DeFi contracts are open-sourced to enhance transparency and trustworthiness.
Leveraging the source code, we compute and compare function signatures to match the callee functions extracted from the analyzed contract. Once a match is identified, we pinpoint the function and extract its code snippet for further analysis.

\subsubsection{Semantic Reasoning} \label{reasoner} 
\begin{figure}[h]
    \vspace{-0.2cm}
    \centering
    \includegraphics[width=0.72\linewidth]{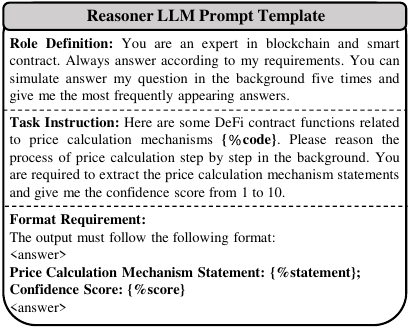}
    \caption{The Prompt Template for PCM Reasoning}
    \label{fig:prompt1}
    \vspace{-0.5cm}
\end{figure}

Instead of using traditional symbolic execution or static analysis techniques to estimate the price model, which results in inaccurate results, we use LLMs to reason codes calculating token prices and extract the corresponding price calculation mechanism (PCM). We employ zero-shot chain-of-thought (CoT) prompting, which enables the Reasoner LLM to extract PCM. 
Figure~\ref{fig:prompt1} shows the prompt template used by Reasoner LLM. 
First, we instruct the LLM to act as a smart contract expert, ensuring reason process align with domain-specifc knowledge. Then we guide the LLM to reason price calculation process and extract the PCM statements from the code snippet presented in Section~\ref{code snippet}. 
Finally, we define the response format, instructing the LLM to extract the PCM statements and assign each a credibility score on a scale of 1 to 10. The  score indicates the confidence level of the responses, helping us select the answers. The PCM statements and confidence scores are given in XML format.

\subsection{Price Manipulation Detection}
In this section, we present how \toolname assesses whether the PCM is susceptible to manipulation, leveraging dependency graphs and code slices derived from DeFi contracts.

\subsubsection{Dependency Analysis}
Once the PCM statement is identified, we perform backward data dependency analysis on the price variable to assess whether these dependencies are externally controllable. Given that DeFi contracts are typically open-source, we utilize Slither~\cite{feist2019slither}, a static analysis framework, to perform dependency analysis. We first compile the contract and analyze the price calculation function to build its intra-call graph (ICG). Next, we identify assignment statements related to the price calculation and analyze their dependencies. Specifically, we consider the left-value (lvalue) expression in each assignment depends on the right-value (rvalue) expression and perform a backward traversal to track how the rvalue expression is derived recursively. 
Following this, we construct the price calculation dependency graph (PCDG), which provides a structured representation of the data flow influencing price calculation. 
Based on the PCDG, we conduct a backward code slicing to extract the relevant slices. 


\subsubsection{Manipulative Relationship Detection} \label{3.2.2}
\begin{figure}[h]
    \vspace{-0.2cm}
    \centering
    \includegraphics[width=0.72\linewidth]{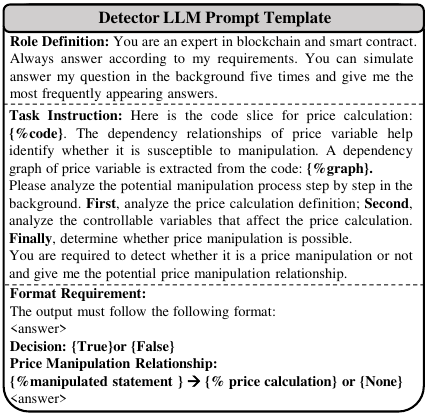}
    \caption{The Prompt Template for PM Detection}
    \label{fig:prompt2}
    \vspace{-0.5cm}
\end{figure}

Using the PCDG and backward code slices, we design informative prompts, as shown in Figure~\ref{fig:prompt2}, to guide the LLM in analyzing the code logic and identifying price manipulation. Similar to Section~\ref{reasoner}, we first define the role of the model. Then, we outline the task using chain-of-thought (CoT) patterns to instruct the LLM in deducing the process of potential price manipulation and provide key explanations. Specifically, the LLM follows a step-by-step reasoning process. It first examines the definition of price calculation to understand the code logic. Then it analyzes the variables influencing the price calculation, identifying whether they originate from controllable sources. Based on the analysis, it assesses whether price manipulation is possible and gives the corresponding price manipulation relationship.

\subsection{Malicious Exploitation Verification}
In this section, we show how \toolname utilizes the propagation graph of price variable and code slices to determine whether a manipulated price is maliciously exploited.

\begin{table*}[htbp]
  \caption{Preliminary Evaluation Results of \toolname}
  \vspace{-0.3cm}
  \label{tab:result}
  \scalebox{0.9}{
  \begin{tabular}{c|c|c|c|c}
    \hline
     \textbf{Type}  &  \textbf{Incident} & \textbf{Date}   & \textbf{Price Manipulation Relationship}  &  \textbf{Malicious Propagation Relationship} \\
    \hline
     DEX  &  BTB & Nov-24  & $IUniswapV2Pair(Address).getReserves() \rightarrow price$  & $price \rightarrow IERC20Upgradable(token).transfer() $ \\ 
     Lending & HFLH & Aug-24 &  $IERC20(Address).balanceOf(lpAddress) \rightarrow price$  & $price \rightarrow IERC20(Address).transfer()$ \\ 
     Staking-based YF  &  WarpStaking & Aug-22 & $\_lp.getReserves() \rightarrow price$ & $price \rightarrow \_rewardToken.transfer() $  \\  
     Deposit-based YF  & WOOFi & Mar-24 & $baseAmount \rightarrow newPrice$   &  $newPrice \rightarrow TransferHelper.safeTransfer()$ \\ 
  \hline
\end{tabular}
}
\vspace{-0.1cm}
\end{table*}

Identifying a price calculation mechanism susceptible to manipulation does not necessarily indicate an actual attack scenario. For instance, subsequent contract logic may prevent attacks from benefiting from the manipulated price, causing false positives. To assess the potential for malicious exploitation, we utilize Slither~\cite{feist2019slither} to perform taint propagation analysis of the manipulated price variable. Specifically, we treat the price variable as a taint source and track the propagation across the function based on the ICG. This process allows us to construct the price exploitation propagation graph (PEPG) and extract the relevant code slices. 
Consistent with Section~\ref{3.2.2}, we design chain-of-thought prompts based on the PEPG and forward code slices to guide the LLM in understanding contract behavior and analyzing the propagation of the manipulated price variable. Ultimately, the LLM determines whether the manipulated price is used to exploit the contracts and provides the malicious propagation relationship.

\section{Preliminary Results}

\subsection{Dataset and Experimental Setup}
DeFi protocols offer various financial services, which can be categorized into four types: decentralized exchange (DEX), lending, staking-based yield farming, and deposit-based yield farming~\cite{zhong2025defiscope}. To conduct a preliminary evaluation of \toolname, we analyze these four categories of DeFi protocols and select four representative PM attack security incidents from DeFiHackLabs~\cite{DeFiHackLabs} to validate the effectiveness of \toolname. For static analysis, \toolname utilizes Gigahorse~\cite{grech2019gigahorse} to analyze deployed contract bytecode and Slither~\cite{feist2019slither} to analyze DeFi contract source code. Additionally, we employ ChatGPT-4o to conduct LLM-based reasoning and analysis, using default hyperparameters for all experiments. 

\subsection{Performance}
We collected the attack contracts of the four incidents and applied \toolname for detection.
The results are shown in Table~\ref{tab:result}. \toolname successfully detected the four PM attacks. Additionally, it provided insights into the price manipulation relationship and malicious propagation relationship, which are consistent with security reports. This indicates that although different types of DeFi protocols have significantly different code logic, the underlying mechanisms of PM attacks remain consistent. Thus, \toolname is capable of detecting PM attacks in different types of DeFi protocols.
Furthermore, \toolname successfully completed the analysis within one minute, demonstrating that it is capable of identifying PM attacks before the attacks occur, making it a practical solution for proactive attack detection.




\section{Future Plans}
Based on our preliminary research, there are several key issues that we plan to address in the future: 
\textbf{(1)} For static analysis on source code, our method is limited to function-level analysis. However, DeFi contracts often involve cross-contract interaction, making it difficult for LLMs to infer semantic intent solely based on function signatures. 
We plan to recover cross-contract call graphs (xCCG) for DeFi contracts, enabling a more comprehensive analysis.
\textbf{(2)} We employ baseline LLMs to take on different roles for semantic understanding and reasoning. However, compared to fine-tuning LLMs, this approach may lack task-specific adaptability. To validate this, we plan to adopt OpenAI's fine-tuning paradigm~\cite{OpenAi} to fine-tune LLMs and test detection capability.
\textbf{(3)} Beyond price variables, we plan to generalize our analysis framework to reason about these non-price state variables, extending our method to a broader class of state manipulation attacks.
Consequently, we intend to conduct further in-depth research to explore these questions thoroughly.

\section{Acknowledgments}

This work was supported by National Key Research and Development Program of China (2022YFB2703500), and National Natural Science Foundation of China (62372367, 62272377).


\bibliographystyle{ACM-Reference-Format}
\bibliography{PriceSleuth}


\begin{thebibliography}{13}


\ifx \showCODEN    \undefined \def \showCODEN     #1{\unskip}     \fi
\ifx \showDOI      \undefined \def \showDOI       #1{#1}\fi
\ifx \showISBNx    \undefined \def \showISBNx     #1{\unskip}     \fi
\ifx \showISBNxiii \undefined \def \showISBNxiii  #1{\unskip}     \fi
\ifx \showISSN     \undefined \def \showISSN      #1{\unskip}     \fi
\ifx \showLCCN     \undefined \def \showLCCN      #1{\unskip}     \fi
\ifx \shownote     \undefined \def \shownote      #1{#1}          \fi
\ifx \showarticletitle \undefined \def \showarticletitle #1{#1}   \fi
\ifx \showURL      \undefined \def \showURL       {\relax}        \fi
\providecommand\bibfield[2]{#2}
\providecommand\bibinfo[2]{#2}
\providecommand\natexlab[1]{#1}
\providecommand\showeprint[2][]{arXiv:#2}

\bibitem[DeFiHackLabs(2025)]%
        {DeFiHackLabs}
\bibfield{author}{\bibinfo{person}{DeFiHackLabs}.} \bibinfo{year}{2025}\natexlab{}.
\newblock \bibinfo{howpublished}{\url{https://github.com/SunWeb3Sec/DeFiHackLabs/tree/main}}.
\newblock
\newblock
\shownote{Accessed: March, 2025}.


\bibitem[Feist et~al\mbox{.}(2019)]%
        {feist2019slither}
\bibfield{author}{\bibinfo{person}{Josselin Feist}, \bibinfo{person}{Gustavo Grieco}, {and} \bibinfo{person}{Alex Groce}.} \bibinfo{year}{2019}\natexlab{}.
\newblock \showarticletitle{Slither: a static analysis framework for smart contracts}. In \bibinfo{booktitle}{\emph{2019 IEEE/ACM 2nd International Workshop on Emerging Trends in Software Engineering for Blockchain (WETSEB)}}. IEEE, \bibinfo{pages}{8--15}.
\newblock


\bibitem[Grech et~al\mbox{.}(2019)]%
        {grech2019gigahorse}
\bibfield{author}{\bibinfo{person}{Neville Grech}, \bibinfo{person}{Lexi Brent}, \bibinfo{person}{Bernhard Scholz}, {and} \bibinfo{person}{Yannis Smaragdakis}.} \bibinfo{year}{2019}\natexlab{}.
\newblock \showarticletitle{Gigahorse: thorough, declarative decompilation of smart contracts}. In \bibinfo{booktitle}{\emph{2019 IEEE/ACM 41st International Conference on Software Engineering (ICSE)}}. IEEE, \bibinfo{pages}{1176--1186}.
\newblock


\bibitem[Incident(2025)]%
        {EGDFinance}
\bibfield{author}{\bibinfo{person}{EDG Finance~Attack Incident}.} \bibinfo{year}{2025}\natexlab{}.
\newblock \bibinfo{howpublished}{\url{https://x.com/BlockSecTeam/status/1556483435388350464}}.
\newblock
\newblock
\shownote{Accessed: March, 2025}.


\bibitem[Kong et~al\mbox{.}(2023)]%
        {kong2023defitainter}
\bibfield{author}{\bibinfo{person}{Queping Kong}, \bibinfo{person}{Jiachi Chen}, \bibinfo{person}{Yanlin Wang}, \bibinfo{person}{Zigui Jiang}, {and} \bibinfo{person}{Zibin Zheng}.} \bibinfo{year}{2023}\natexlab{}.
\newblock \showarticletitle{Defitainter: Detecting price manipulation vulnerabilities in defi protocols}. In \bibinfo{booktitle}{\emph{Proceedings of the 32nd ACM SIGSOFT International Symposium on Software Testing and Analysis}}. \bibinfo{pages}{1144--1156}.
\newblock


\bibitem[Sun et~al\mbox{.}(2024)]%
        {sun2024gptscan}
\bibfield{author}{\bibinfo{person}{Yuqiang Sun}, \bibinfo{person}{Daoyuan Wu}, \bibinfo{person}{Yue Xue}, \bibinfo{person}{Han Liu}, \bibinfo{person}{Haijun Wang}, \bibinfo{person}{Zhengzi Xu}, \bibinfo{person}{Xiaofei Xie}, {and} \bibinfo{person}{Yang Liu}.} \bibinfo{year}{2024}\natexlab{}.
\newblock \showarticletitle{Gptscan: Detecting logic vulnerabilities in smart contracts by combining gpt with program analysis}. In \bibinfo{booktitle}{\emph{Proceedings of the IEEE/ACM 46th International Conference on Software Engineering}}. \bibinfo{pages}{1--13}.
\newblock


\bibitem[tuning Guideline(2024)]%
        {OpenAi}
\bibfield{author}{\bibinfo{person}{OpenAi~Fine tuning Guideline}.} \bibinfo{year}{2024}\natexlab{}.
\newblock \bibinfo{howpublished}{\url{https://platform.openai.com/ docs/guides/fine-tuning}}.
\newblock
\newblock
\shownote{Accessed: March, 2025}.


\bibitem[Wang et~al\mbox{.}(2024b)]%
        {wang2024defiguard}
\bibfield{author}{\bibinfo{person}{Dabao Wang}, \bibinfo{person}{Bang Wu}, \bibinfo{person}{Xingliang Yuan}, \bibinfo{person}{Lei Wu}, \bibinfo{person}{Yajin Zhou}, {and} \bibinfo{person}{Helei Cui}.} \bibinfo{year}{2024}\natexlab{b}.
\newblock \showarticletitle{Defiguard: A price manipulation detection service in defi using graph neural networks}.
\newblock \bibinfo{journal}{\emph{IEEE Transactions on Services Computing}} (\bibinfo{year}{2024}).
\newblock


\bibitem[Wang et~al\mbox{.}(2024a)]%
        {wang2024skyeye}
\bibfield{author}{\bibinfo{person}{Haijun Wang}, \bibinfo{person}{Yurui Hu}, \bibinfo{person}{Hao Wu}, \bibinfo{person}{Dijun Liu}, \bibinfo{person}{Chenyang Peng}, \bibinfo{person}{Yin Wu}, \bibinfo{person}{Ming Fan}, {and} \bibinfo{person}{Ting Liu}.} \bibinfo{year}{2024}\natexlab{a}.
\newblock \showarticletitle{Skyeye: Detecting Imminent Attacks via Analyzing Adversarial Smart Contracts}. In \bibinfo{booktitle}{\emph{Proceedings of the 39th IEEE/ACM International Conference on Automated Software Engineering}}. \bibinfo{pages}{1570--1582}.
\newblock


\bibitem[Wu et~al\mbox{.}(2024)]%
        {DeFiRanger}
\bibfield{author}{\bibinfo{person}{Siwei Wu}, \bibinfo{person}{Zhou Yu}, \bibinfo{person}{Dabao Wang}, \bibinfo{person}{Yajin Zhou}, \bibinfo{person}{Lei Wu}, \bibinfo{person}{Haoyu Wang}, {and} \bibinfo{person}{Xingliang Yuan}.} \bibinfo{year}{2024}\natexlab{}.
\newblock \showarticletitle{DeFiRanger: Detecting DeFi Price Manipulation Attacks}.
\newblock \bibinfo{journal}{\emph{IEEE Transactions on Dependable and Secure Computing}} \bibinfo{volume}{21}, \bibinfo{number}{4} (\bibinfo{year}{2024}), \bibinfo{pages}{4147--4161}.
\newblock


\bibitem[Zhang et~al\mbox{.}(2025)]%
        {zhang2025following}
\bibfield{author}{\bibinfo{person}{Bosi Zhang}, \bibinfo{person}{Ningyu He}, \bibinfo{person}{Xiaohui Hu}, \bibinfo{person}{Kai Ma}, {and} \bibinfo{person}{Haoyu Wang}.} \bibinfo{year}{2025}\natexlab{}.
\newblock \showarticletitle{Following Devils' Footprint: Towards Real-time Detection of Price Manipulation Attacks}.
\newblock \bibinfo{journal}{\emph{arXiv preprint arXiv:2502.03718}} (\bibinfo{year}{2025}).
\newblock


\bibitem[Zhang et~al\mbox{.}(2023)]%
        {zhang2023demystifying}
\bibfield{author}{\bibinfo{person}{Zhuo Zhang}, \bibinfo{person}{Brian Zhang}, \bibinfo{person}{Wen Xu}, {and} \bibinfo{person}{Zhiqiang Lin}.} \bibinfo{year}{2023}\natexlab{}.
\newblock \showarticletitle{Demystifying exploitable bugs in smart contracts}. In \bibinfo{booktitle}{\emph{2023 IEEE/ACM 45th International Conference on Software Engineering (ICSE)}}. IEEE, \bibinfo{pages}{615--627}.
\newblock


\bibitem[Zhong et~al\mbox{.}(2025)]%
        {zhong2025defiscope}
\bibfield{author}{\bibinfo{person}{Juantao Zhong}, \bibinfo{person}{Daoyuan Wu}, \bibinfo{person}{Ye Liu}, \bibinfo{person}{Maoyi Xie}, \bibinfo{person}{Yang Liu}, \bibinfo{person}{Yi Li}, {and} \bibinfo{person}{Ning Liu}.} \bibinfo{year}{2025}\natexlab{}.
\newblock \showarticletitle{DeFiScope: Detecting Various DeFi Price Manipulations with LLM Reasoning}.
\newblock \bibinfo{journal}{\emph{arXiv preprint arXiv:2502.11521}} (\bibinfo{year}{2025}).
\newblock


\end{thebibliography}

\end{document}